\documentclass[prl,aps,showkeys,showpacs,superscriptaddress,twocolumn,%
amsmath,amssymb]{revtex4} 

\usepackage{graphics}
\usepackage{epsfig}
\usepackage{epstopdf}
\usepackage{amsmath}
\usepackage{setspace}
\usepackage[bookmarks=false,linkcolor=blue,urlcolor=black,colorlinks,citecolor=blue]{hyperref}

\begin{document}
\title{Comment on ``Field Theory for Amorphous Solids"  by Eric DeGiuli}
\author{Bulbul Chakraborty}
\email{bulbul@brandeis.edu}
\affiliation{Martin Fisher School of Physics, Brandeis University, Waltham, MA 02454, USA}
\author{Silke Henkes}
\email{silke.henkes@bristol.ac.uk}
\affiliation{Department of Mathematics, University of Bristol, Bristol BS8 1TH, UK}

\date{\today} 


\begin{abstract}
In a recent pair of papers, Eric DeGiuli has developed a field theory of glasses and granular materials based on the Edwards ensemble, extending our earlier theoretical framework.  In this comment,  we address a misconception regarding the relation between equiprobability of microstates and a flat measure in a field theory, which appears in the DeGiuli papers, and has often plagued discussions surrounding the Edwards ensemble.  We  point out that modeling this measure is the challenge addressed in both our earlier work and the recent DeGiuli work.  Contrary to what is stated in the the DeGiuli papers, we did not assume a flat measure in our earlier work.
\end{abstract}

\maketitle

In a recent pair of papers \cite{eric_prl,eric_pre}, Eric DeGiuli has presented a field theory of glasses and granular materials based on the Edwards ensemble \cite{Edwards,ann_rev_bi}.  A major result to emerge from this theory is that long-range stress correlations are a consequence of the constraints of mechanical equilibrium \cite{eric_prl,eric_pre,delgado_viewpoint}.  As pointed out by the author \cite{eric_prl,eric_pre}, the theoretical framework extends our earlier work \cite{henkes_2005,henkes_2009}, and the results DeGiuli obtains in the Gaussian approximation are exactly the same as our earlier results, including the long-range correlations of the stress \cite{henkes_lois_pre_rapid_2009}.   

The field theory presented by DeGiuli could very well become the general elasticity theory of amorphous solids, and thus it promises to have a significant impact on the field of disordered, glassy systems.  It is, therefore, essential to clarify the assumptions that have entered the theory and the construction of the field theory.  Unfortunately, a statement early on in the paper propagates a misconception that has often plagued discussions of the Edwards ensemble, and is an incorrect representation of our earlier work.  
To be clear, although the statement is wrong, it does not affect the results presented in the paper.  However, it is a misrepresentation of which assumptions have gone into the current theory, and how it compares to our  earlier work. Therefore, we feel compelled to correct and clarify this misconception.

The statement appears after Eq. 10 in \cite{eric_prl} and Eq. 15 in \cite{eric_pre}.  To quote: ``In a strict canonical ensemble, the sampling probability
$\omega$ would be unity, as was taken in previous work on the stress ensemble [28,42]. In fact, there is no general
justification for the flat measure out of equilibrium, even if it was observed to hold to a good approximation in several model systems [44,45].''  

This statement as written is misleading,  and is a misrepresentation of our work (Refs 28 and 42 above), as we did not assume $\omega$ to be unity.    $\omega[\hat{\sigma}[\psi]]$ is a measure of the density (number) of microstates that are consistent with a given field configuration $\psi(\bf r)$, the Airy stress tensor field.   It is thus the sampling probability of the continuum field, {\it and not of microstates}.   There is no justification for taking this to be a flat measure either in or out of equilibrium.

The analog of $\omega$ in more familiar equilibrium statistical mechanics models such as the Ising model is $\Omega[m(\bf r)]$, which is the number of microscopic spin configurations that are consistent with a coarse-grained magnetization field, $m(\bf r)$.  In equilibrium statistical mechanics, all microstates with the same energy (the microcanonical ensemble) are equiprobable.  However, {\it the density of states, $\Omega[m(\bf r)]$ is manifestly not  independent of $m(\bf r)$ except for a non interacting system}.   Equiprobability of microstates does not imply a flat sampling probability of the continuum fields in a canonical ensemble.  The challenge  and the central problem faced in constructing a statistical field theory is to model this density of states, or get exact expressions in special cases (exactly solvable models).   The Boltzmann weight needed to go from the microcanonical to the canonical ensemble is simply written down once the energy function is specified.  

The canonical Edwards stress ensemble has an analog of the Boltzmann weight, as discussed in DeGiuli's papers; however to complete the field theory, one still has to construct a model for $\omega[\hat \sigma[\psi(\bf r)]]$, the analog of $\Omega[m(\bf r)]$.    The flat measure that DeGiuli refers to in the context of the Edwards ensemble is the weight of each microstate in the microcanonical ensemble.  In equilibrium statistical mechanics, $\Omega[m(\bf r)] = \sum_\nu \delta(m(\bf r) - m_\nu)$, where $\nu$ denotes a microstates and equiprobability has been assumed.  As we have discussed in the context of the Edwards ensemble, if equiprobability is violated, we can still construct an analog of the density of states under a set of weaker conditions \cite{ann_rev_bi}.  Postulating an Edwards ensemble means we assume that these weaker conditions hold.
If the microstates have different weights $\gamma_\nu$, then the definition of the generalized density of states is $\omega[\hat \sigma[\psi(\bf r)]]$ is $ \omega[\hat \sigma(\bf r)] =\sum_\nu \gamma_\nu \delta(\hat \sigma(\bf r) - \hat \sigma_\nu)$. 
 
 Modeling $ \omega[\hat \sigma(\bf r)]$ is the crux of constructing a Edwards field theory of amorphous materials, and is at the heart of both the DeGiuli papers and our earlier work. Both approaches are based on coarse-graining, which is not possible to do explicitly for any but the simplest systems. We, therefore, construct the simplest models consistent with symmetries and constraints.  There is no conceivable way, in this modeling, to decouple the effect of $\gamma_\nu$ and the counting represented by $\delta(\hat \sigma(\bf r) - \hat \sigma_\nu)$.   In our earlier work we constructed an expression for $\omega[\hat \sigma[\psi(\bf r)]]$.  {\it We did not assume that it was unity}.    In fact our expression is rather similar to that in Ref. \cite{eric_prl,eric_pre}, which is why DeGiuli's results in the Gaussian approximation coincides with ours.

The DeGiuli papers do not shed any light on the validity of the equiprobability assumption of Edwards \cite{Edwards}.  The Edwards field theory does assume the weaker condition of factorizability of states: $\gamma_{\nu + \nu^\prime} = \gamma_\nu \gamma_\nu^\prime$, which is necessary for the validity of the canonical weights in Eq. 9 (Ref.\cite{eric_prl}).   The long-range stress correlations are a consequence of constraints of mechanical equilibrium, which in turn strongly constrain the form of $\omega[\hat \sigma[\psi(\bf r)]]$.  The models proposed for $\omega[\hat \sigma[\psi(\bf r)]]$ obey these constraints \cite{eric_prl,eric_pre,henkes_2009,henkes_lois_pre_rapid_2009}, but are agnostic about Edwards equiprobability.   

\bibliography{comment_prl} 
\bibliographystyle{apsrev4-1}

\end{document}